\documentclass[
    ,final            
  ]
  {aipproc}

\layoutstyle{8x11single}

\newcommand{\MJ}{\textsc{Majo\-ra\-na}}
\newcommand{\DEM}{\textsc{Dem\-on\-strat\-or}}

\newcommand{\MAGE}{\textsc{MaGe}}

\newcommand{\nonubb}{$0 \nu \beta \beta$}

\def\nuc#1#2{${}^{\mathrm{#1}}$#2}

\begin{document}

\title{The \MJ\ \DEM: A Search for Neutrinoless Double-beta Decay of
Germanium-76}



\newcommand{\alberta}{Centre for Particle Physics, University of Alberta, Edmonton, AB, Canada}
\newcommand{\blhill}{Department of Physics, Black Hills State University, Spearfish, SD, USA}
\newcommand{\ITEP}{Institute for Theoretical and Experimental Physics, Moscow, Russia}
\newcommand{\JINR}{Joint Institute for Nuclear Research, Dubna, Russia}
\newcommand{\lbnl}{Nuclear Science Division, Lawrence Berkeley National Laboratory, Berkeley, CA, USA}
\newcommand{\lanl}{Los Alamos National Laboratory, Los Alamos, NM, USA}
\newcommand{\queens}{Department of Physics, Queen's University, 
Kingston, ON, Canada}
\newcommand{\uw}{Center for Experimental Nuclear Physics and Astrophysics 
and Department of Physics, University of Washington, Seattle, WA, USA}
\newcommand{\uchic}{Department of Physics, University of Chicago, Chicago, IL, USA}
\newcommand{\unc}{Department of Physics and Astronomy, University of North Carolina, Chapel Hill, NC, USA}
\newcommand{\duke}{Department of Physics, Duke University, Durham, NC, USA}
\newcommand{\ncsu}{Department of Physics, North Carolina State University, Raleigh, NC, USA}
\newcommand{\ornl}{Oak Ridge National Laboratory, Oak Ridge, TN, USA}
\newcommand{\ou}{Research Center for Nuclear Physics and Department of Physics, Osaka University, Ibaraki, Osaka, Japan}
\newcommand{\pnnl}{Pacific Northwest National Laboratory, Richland, WA, USA}
\newcommand{\sdsmt}{South Dakota School of Mines and Technology, Rapid City, SD, USA}
\newcommand{\usc}{Department of Physics and Astronomy, University of South Carolina, Columbia, SC, USA}
\newcommand{\usd}{Department of Earth Science and Physics, University of South Dakota, Vermillion, SD, USA}
\newcommand{\ut}{Department of Physics and Astronomy, University of Tennessee, Knoxville, TN, USA}
\newcommand{\tunl}{Triangle Universities Nuclear Laboratory, Durham, NC, USA}


\author{A.G.~Schubert}{
  address={\uw},
}
\author{E.~Aguayo}{address=\pnnl} 
\author{F.T.~Avignone~III}{address=\usc, altaddress=\ornl}
\author{H.O.~Back}{address=\ncsu, altaddress=\tunl} 
\author{A.S.~Barabash}{address=\ITEP}
\author{M.~Bergevin}{address=\lbnl} 
\author{F.E.~Bertrand}{address=\ornl}
\author{M.~Boswell}{address=\lanl} 
\author{V.~Brudanin}{address=\JINR}
\author{M.~Busch}{address=\duke, altaddress=\tunl}	
\author{Y-D.~Chan}{address=\lbnl}
\author{C.D.~Christofferson}{address=\sdsmt} 
\author{J.I.~Collar}{address=\uchic}
\author{D.C.~Combs}{address=\ncsu, altaddress=\tunl}  
\author{R.J.~Cooper}{address=\ornl}
\author{J.A.~Detwiler}{address=\lbnl}
\author{J.~Leon}{address=\uw}	
\author{P.J.~Doe}{address=\uw}

\author{Yu.~Efremenko}{address=\ut}
\author{V.~Egorov}{address=\JINR}
\author{H.~Ejiri}{address=\ou}
\author{S.R.~Elliott}{address=\lanl}
\author{J.~Esterline}{address=\duke, altaddress=\tunl}
\author{J.E.~Fast}{address=\pnnl}
\author{N.~Fields}{address=\uchic} 
\author{P.~Finnerty}{address=\unc, altaddress=\tunl}
\author{F.M.~Fraenkle}{address=\unc, altaddress=\tunl} 
\author{V.M.~Gehman}{address=\lanl}
\author{G.K.~Giovanetti}{address=\unc, altaddress=\tunl}  
\author{M.P.~Green}{address=\unc, altaddress=\tunl}  
\author{V.E.~Guiseppe}{address=\usd}	
\author{K.~Gusey}{address=\JINR}
\author{A.L.~Hallin}{address=\alberta}
\author{R.~Hazama}{address=\ou}
\author{R.~Henning}{address=\unc, altaddress=\tunl}
\author{A.~Hime}{address=\lanl}
\author{E.W.~Hoppe}{address=\pnnl}
\author{M.~Horton}{address=\sdsmt} 
\author{S.~Howard}{address=\sdsmt}  
\author{M.A.~Howe}{address=\unc, altaddress=\tunl}

\author{R.A.~Johnson}{address=\uw} 
\author{K.J.~Keeter}{address=\blhill}
\author{M.E.~Keillor}{address=\pnnl}
\author{C.~Keller}{address=\usd}
\author{J.D.~Kephart}{address=\pnnl} 
\author{M.F.~Kidd}{address=\lanl}	
\author{A. Knecht}{address=\uw}	
\author{O.~Kochetov}{address=\JINR}
\author{S.I.~Konovalov}{address=\ITEP}
\author{R.T.~Kouzes}{address=\pnnl}
\author{B.~LaFerriere}{address=\pnnl}
\author{B.H.~LaRoque}{address=\lanl}	
\author{L.E.~Leviner}{address=\ncsu, altaddress=\tunl}
\author{J.C.~Loach}{address=\lbnl}	
\author{S.~MacMullin}{address=\unc, altaddress=\tunl}
\author{M.G.~Marino}{address=\uw}
\author{R.D.~Martin}{address=\lbnl}	
\author{D.-M.~Mei}{address=\usd}
\author{J.~Merriman}{address=\pnnl}
\author{M.L.~Miller}{address=\uw} 
\author{L.~Mizouni}{address=\usc, altaddress=\pnnl}  
\author{M.~Nomachi}{address=\ou}
\author{J.L.~Orrell}{address=\pnnl}
\author{N.~Overman}{address=\pnnl}
\author{D.G.~Phillips~II}{address=\unc, altaddress=\tunl}  
\author{A.W.P.~Poon}{address=\lbnl}
\author{G.~Perumpilly}{address=\usd}   
\author{G.~Prior}{address=\lbnl} 
\author{D.C.~Radford}{address=\ornl}
\author{K.~Rielage}{address=\lanl}
\author{R.G.H.~Robertson}{address=\uw}
\author{M.C.~Ronquest}{address=\lanl}	
\author{T.~Shima}{address=\ou}
\author{M.~Shirchenko}{address=\JINR}
\author{K.J.~Snavely}{address=\unc, altaddress=\tunl}	
\author{V.~Sobolev}{address=\sdsmt}  
\author{D.~Steele}{address=\lanl}	
\author{J.~Strain}{address=\unc, altaddress=\tunl}
\author{K.~Thomas}{address=\usd}		
\author{V.~Timkin}{address=\JINR}
\author{W.~Tornow}{address=\duke, altaddress=\tunl}
\author{I.~Vanyushin}{address=\ITEP}
\author{R.L.~Varner}{address=\ornl}  
\author{K.~Vetter}{address={Alternate address: Department of Nuclear
Engineering, University of California, Berkeley, CA, USA}, altaddress=\lbnl}
\author{K.~Vorren}{address=\unc, altaddress=\tunl} 
\author{J.F.~Wilkerson}{address=\unc, altaddress=\tunl, altaddress=\ornl}    
\author{B.A.~Wolfe}{address=\uw}	
\author{E.~Yakushev}{address=\JINR}
\author{A.R.~Young}{address=\ncsu, altaddress=\tunl}
\author{C.-H.~Yu}{address=\ornl}
\author{V.~Yumatov}{address=\ITEP}
\author{C.~Zhang}{address=\usd}

\begin{abstract}

The observation of neutrinoless double-beta decay would determine whether the
neutrino is a Majorana particle and provide information on the absolute scale of
neutrino mass. The \MJ\ Collaboration is constructing the \DEM, an array of
germanium detectors, to search for neutrinoless double-beta decay of
\nuc{76}{Ge}.  The \DEM\ will contain 40 kg of germanium; up to 30 kg will be
enriched to 86\% in \nuc{76}{Ge}.  The \DEM\ will be deployed deep underground
in an ultra-low-background shielded environment.  Operation of the \DEM\ aims to
determine whether a future tonne-scale germanium experiment can achieve a
background goal of one count per tonne-year in a 4-keV region of interest around
the \nuc{76}{Ge} neutrinoless double-beta decay Q-value of 2039 keV.  

\end{abstract}

\maketitle


\section{Introduction}

Observation of neutrinoless double-beta decay (\nonubb) would determine the
Majorana nature of the neutrino and would demonstrate that lepton number is not
conserved.  Additionally, the measurement of a \nonubb\ rate will provide
information about the absolute scale of neutrino mass.  The \MJ\
Collaboration~\cite{Aalseth201144} will search for neutrinoless double beta
decay with an array of germanium detectors enriched in the \nonubb\ candidate
isotope \nuc{76}{Ge}.

The \MJ\ and GERDA~\cite{Gerda} Collaborations are independently investigating
technologies for deployment of ultra-low-background germanium detector arrays as
research and development toward a future collaborative tonne-scale \nonubb\
experiment.  As part of this effort, \MJ\ is constructing the \DEM, a germanium
detector array, with the goal of testing a recent claimed observation of
\nonubb~\cite{KlapdorKleingrothaus:2006ff} and determining the background rates
achievable with the proposed compact shield design.  For a tonne-scale
experiment, \MJ\ and GERDA have a background goal of one count per tonne-year
after analysis cuts in a narrow energy region surrounding the double-beta decay
endpoint at 2039 keV.  The energy region of interest (ROI) is expected to be
approximately 4 keV wide, with the actual width depending on detector
resolution.  This background goal is approximately 100 times more stringent than
background rates observed in previous low-background germanium
experiments~\cite{HM}, \cite{IGEX}.

\section{The \MJ\ \DEM}

The \DEM\ will contain 40 kg of germanium semiconductor diode detectors, of
which at least 20 kg and up to 30 kg will be enriched to 86\% in \nuc{76}{Ge}.
The \DEM\ will be located at the 4850-foot level of Sanford Underground Research
Facility in Lead, SD.  An array of enriched detectors is expected to begin
operation in 2013.  Germanium detectors will be housed in two vacuum cryostats
constructed from electroformed copper and cooled with liquid nitrogen.  The
detectors will be mounted in \textit{strings}, rigid columns of detectors
suspended from an electroformed copper cold plate.  Each cryostat will contain
seven strings of five detectors in a close-packed geometry.  

The \DEM\ cryostats will be surrounded by several layers of shielding, shown in
Fig.~1.  In order of proximity to the detectors, these layers will consist of
electroformed copper, commercial copper, lead, a radon exclusion volume, an
active muon scintillator veto, and polyethylene neutron moderator.  A background
count rate at or below four counts per tonne-year in the \DEM\ ROI would scale
to the background goal of one count per tonne-year in the ROI in a tonne-scale
\nuc{76}{Ge} experiment.

\begin{figure}[h]
\includegraphics[height=0.18\textheight]{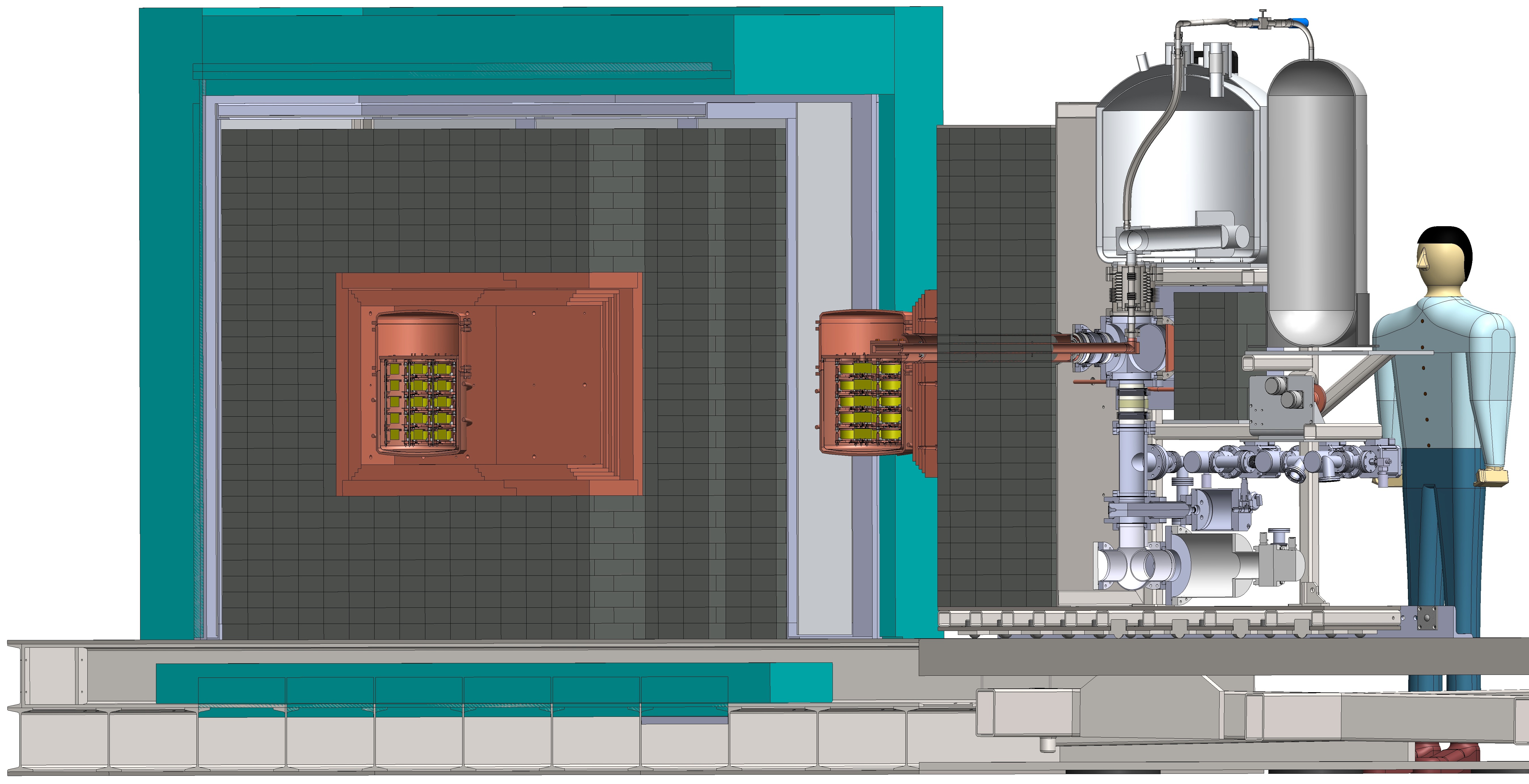}
\caption{%
A drawing of the \MJ\ \DEM, showing insertion of a cryostat module
into the shield.
}
\end{figure}

\section{\nonubb\ detection and background identification techniques}

A \nonubb\ signal would appear as a peak at the Q-value in the energy spectrum
collected by the \DEM.  The sensitivity of the \DEM\ is limited by background
contributions in the ROI; any radiation with energy greater than the Q-value
could potentially create a background in the ROI.  
Backgrounds are expected from in-situ cosmogenic flux, long-lived unstable
cosmogenically activated contaminants such as \nuc{68}{Ge} in the detectors and
\nuc{60}{Co} in the copper, and primordial radioimpurities, such as \nuc{40}{K},
\nuc{238}{U}, and \nuc{232}{Th} decay products.  
Only materials certified by ongoing assay and simulation campaigns will be used.
Component mass is minimized to reduce contaminants.  Close packing of the
detectors enhances effectiveness of a crystal-to-crystal \textit{granularity}
tag.  
Measurements in dedicated thermal and electronics test cryostats have refined
the designs of front-end electronics and detector mounts.

\MJ\ has chosen p-type point contact (PPC) germanium detectors for their
superior background rejection capabilities.  The low energy threshold achievable
with these detectors should allow identification of certain backgrounds that are
time-correlated with low-energy events and will provide sensitivity to some dark
matter candidates.  
Enriched germanium detectors provide an advantageous source-as-detector
configuration with extremely high efficiency for detecting potential \nonubb\
decays.  Germanium detectors exhibit excellent energy resolution,
approximately 0.2\% FWHM at the \nonubb\ Q-value, which allows the definition of
a narrow ROI and minimizes contributions of backgrounds.  

Pulse-shape analysis can distinguish an energy deposit occurring in a single
site in a detector, characteristic of the \nonubb\ signal, from multi-site
interactions characteristic of gamma backgrounds.  This distinction can be made
with with high efficiency for accepting single-site events and rejecting
backgrounds~\cite{Cooper2011303}.  A gamma ray from a background source may
deposit energy in one detector, exit the detector, and deposit energy in another
detector.  These background events can be identified with the granularity tag.
Some radioactive decays, including the decay of \nuc{68}{Ge} to \nuc{68}{Ga},
may be identified not only by their multiple-site nature, but also by analysis
of correlations in time, energy, and location.  

Ultra-clean fabrication efforts are producing components for the \DEM.
Copper electroforming is underway at underground laboratories at a shallow site
at Pacific Northwest National Laboratory in Richland, WA, and at the 4850-foot
level of Sanford Underground Research Facility.  
A prototype Parylene deposition system has produced low-background
Parylene-coated copper ribbon cable for readout of detector electronics.

Operating for five years, the \DEM\ will achieve only a fraction of a tonne-year
of exposure, and less than one count is expected in the ROI.  A thorough
understanding of the full \DEM\ background energy spectrum will be required to
predict the background count rate in the ROI of a tonne-scale \nuc{76}{Ge}
experiment.  By measuring the energy spectrum in a window up to 250 keV wide
surrounding the \nonubb\ Q-value, a sensitive measurement can be made with \DEM\
and extrapolated to the tonne scale.  
The Monte-Carlo physics simulation package \MAGE~\cite{MaGe2011} is used to
simulate \DEM\ detector response to backgrounds.
\MAGE\ results are combined with information from the material assay campaign to
predict the \DEM\ background energy spectrum.

\section{Conclusions}

\MJ\ is constructing the \DEM, a 40-kg germanium detector array to search for
\nonubb\ of \nuc{76}{Ge}.  Efforts in development of ultra-low-background
electronics and cabling, fabrication of radiopure components, material assay,
and modeling of backgrounds are underway.  The \DEM\ should confirm or refute a
previous claim of \nonubb\ observation and will determine whether a background
goal of one count per tonne-year is achievable for a tonne-scale experiment.
The \DEM\ is expected to operate with enriched detectors in 2013.




\bibliographystyle{aipproc}   

\bibliography{4E3_Schubert}

\end{document}